\documentclass[aps,prl,onecolumn,showpacs,amsmath,amssymb,superscriptaddress]{revtex4}
\usepackage{graphicx}
\usepackage{amssymb}
\usepackage{natbib}

\begin{document}

\title{Uniaxial pressure effect on structural and magnetic phase transitions in 
NaFeAs and its comparison with as-grown and annealed BaFe$_2$As$_2$ 
}
\author{Yu Song}
\affiliation{ Department of Physics and Astronomy,
The University of Tennessee, Knoxville, Tennessee 37996-1200, USA }
\author{Scott V. Carr}
\affiliation{ Department of Physics and Astronomy,
The University of Tennessee, Knoxville, Tennessee 37996-1200, USA }
\author{Xingye Lu}
\affiliation{Beijing National Laboratory for Condensed Matter
Physics, Institute of Physics, Chinese Academy of Sciences, Beijing
100190, China}
\affiliation{ Department of Physics and Astronomy,
The University of Tennessee, Knoxville, Tennessee 37996-1200, USA }
\author{Chenglin Zhang}
\affiliation{ Department of Physics and Astronomy,
The University of Tennessee, Knoxville, Tennessee 37996-1200, USA }
\author{Zachary C. Sims}
\affiliation{ Department of Physics and Astronomy,
The University of Tennessee, Knoxville, Tennessee 37996-1200, USA }
\author{N. F. Luttrell}
\affiliation{ Department of Physics and Astronomy,
The University of Tennessee, Knoxville, Tennessee 37996-1200, USA }
\author{Songxue Chi}
\affiliation{Quantum Condensed Matter Division, Oak Ridge National Laboratory, Oak Ridge, Tennessee 37831, USA
}
\author{Yang Zhao}
\affiliation{NIST Center for Neutron Research, National Institute of Standards and Technology, Gaithersburg, Maryland 20899, USA
}
\affiliation{
Department of Materials Science and Engineering, University of Maryland, College Park, Maryland 20742, USA
}
\author{Jeffrey W. Lynn}
\affiliation{NIST Center for Neutron Research, National Institute of Standards and Technology, Gaithersburg, Maryland 20899, USA
}
\author{Pengcheng Dai}
\email{pdai@utk.edu}
\affiliation{ Department of Physics and Astronomy,
The University of Tennessee, Knoxville, Tennessee 37996-1200, USA }
\affiliation{Beijing National Laboratory for
Condensed Matter Physics, Institute of Physics, Chinese Academy of
Sciences, Beijing 100190, China}

\date{\today}
\pacs{74.70.Xa, 74.62.Fj, 75.40.Cx, 75.50.Ee}

\begin{abstract}
We use neutron scattering to study the effect of uniaxial pressure on 
the tetragonal-to-orthorhombic structural ($T_s$) and paramagnetic-to-antiferromagnetic ($T_N$) 
phase transitions in  
NaFeAs and compare the outcome with similar measurements on as-grown and annealed BaFe$_2$As$_2$.
In previous work on as-grown BaFe$_2$As$_2$, uniaxial pressure necessary to detwin the sample was found to induce a significant increase in zero pressure $T_N$ and $T_s$. However, we find that 
similar uniaxial pressure used to detwin NaFeAs and annealed BaFe$_2$As$_2$
 has a very small effect on their $T_N$ and $T_s$.
Since transport measurements on these samples still reveal resistivity anisotropy above $T_N$ and $T_s$,
we conclude that such anisotropy cannot be due to uniaxial strain induced $T_N$ and $T_s$ shifts,
but must arise from intrinsic electronic anisotropy in these materials.
\end{abstract}

\maketitle

The parent compounds of iron pnictide superconductors such as NaFeAs and BaFe$_2$As$_2$
exhibit a tetragonal-to-orthorhombic lattice distortion at temperature $T_s$ and 
paramagnetic-to-antiferromagnetic phase transition at $T_N$ ($\leq T_s$), forming a low-temperature collinear antiferromagnetic (AF) state with  
ordering wave vector along the $[\pm 1,0]$ directions of the orthorhombic lattice [Figs. 1(a) and 1(b) ] \cite{kamihara,cruz,cwchu,slli,drparker,qhunag,mgkim}.
Because of the twinning effect in the orthorhombic AF state, AF Bragg peaks from the twinned domains in Fig. 1(c)
should occur at $[\pm 1,0]$ and $[0,\pm 1]$ positions in reciprocal space [Fig. 1(d)] \cite{dai}. 
To probe the possible electronic anisotropic state (the electronic nematic phase) 
that breaks the $C_4$ rotational symmetry of the paramagnetic tetragonal phase in iron pnictides \cite{fradkin}, 
one needs to prepare single domain samples by applying a uniaxial pressure (strain)
along one-axis of the orthorhombic lattice \cite{jhchu,matanatar}.  Indeed, transport measurements on uniaxial pressure
detwinned samples of NaFeAs \cite{yzhang} and BaFe$_2$As$_2$ \cite{fisher} reveal clear resistivity anisotropy above the zero pressure $T_N$ and $T_s$ that has been interpreted as arising 
from the spin nematic phase \cite{jphu,kasahara,fernandes10}  or orbital ordering  \cite{myi,cclee,kruger,lv,mdaghofer,ccchen,valenzeula} 
in the paramagnetic tetragonal state.  However, recent neutron scattering experiments on as-grown BaFe$_2$As$_2$ find that
a uniaxial pressure necessary to detwin the sample can also induce a significant ($\sim$10 K) 
upward shift in $T_N$ and $T_s$ \cite{dhital}, 
suggesting that the observed resistivity anisotropy above the stress-free $T_N$ and $T_s$ in detwinned samples \cite{yzhang,fisher} may
actually occur in the AF ordered orthorhombic state below the strain-induced $T_N$ and $T_s$.  Furthermore, the resistivity anisotropy above $T_N$ and $T_s$ in as-grown BaFe$_2$As$_2$ 
and electron-doped BaFe$_{2-x}$Co$_x$As$_2$ 
becomes much smaller in annealed samples \cite{nakajima,ishida}, suggesting that the observed resistivity anisotropy in the tetragonal phase 
is not intrinsic to these materials but arises from the anisotropic impurity scattering of Co-atoms in the
FeAs layer \cite{ishida,allan}.

Here we use neutron scattering to study the uniaxial pressure effect on magnetic
and structural phase transitions in NaFeAs \cite{slli}, as-grown, and 
annealed BaFe$_2$As$_2$ \cite{ishida}.  While our measurements on as-grown 
BaFe$_2$As$_2$ confirm the earlier work that the uniaxial pressure necessary to detwin the crystal 
also causes significant increases in $T_N$ and $T_s$ \cite{dhital}, we find that similar uniaxial pressure has
a very small effect on the magnetic and structural phase transitions in NaFeAs and annealed BaFe$_2$As$_2$.
Since transport measurements on identical NaFeAs and annealed BaFe$_2$As$_2$ show clear resistivity anisotropy at temperatures well above the $T_N$ and $T_s$ under uniaxial pressure, we conclude that 
the resistivity anisotropy seen in detwinned NaFeAs and annealed BaFe$_2$As$_2$ in the paramagnetic tetragonal phase must be intrinsic properties of these materials, and cannot be due to the impurity 
scattering effect \cite{ishida,allan}.  These results suggest the presence of an electronic nematic state in the paramagnetic tetragonal 
phase unrelated to the Co-impurity scattering in electron-doped BaFe$_{2-x}$Co$_x$As$_2$ family of materials \cite{ishida,allan}.

\begin{figure}[t]
\includegraphics[scale=.25]{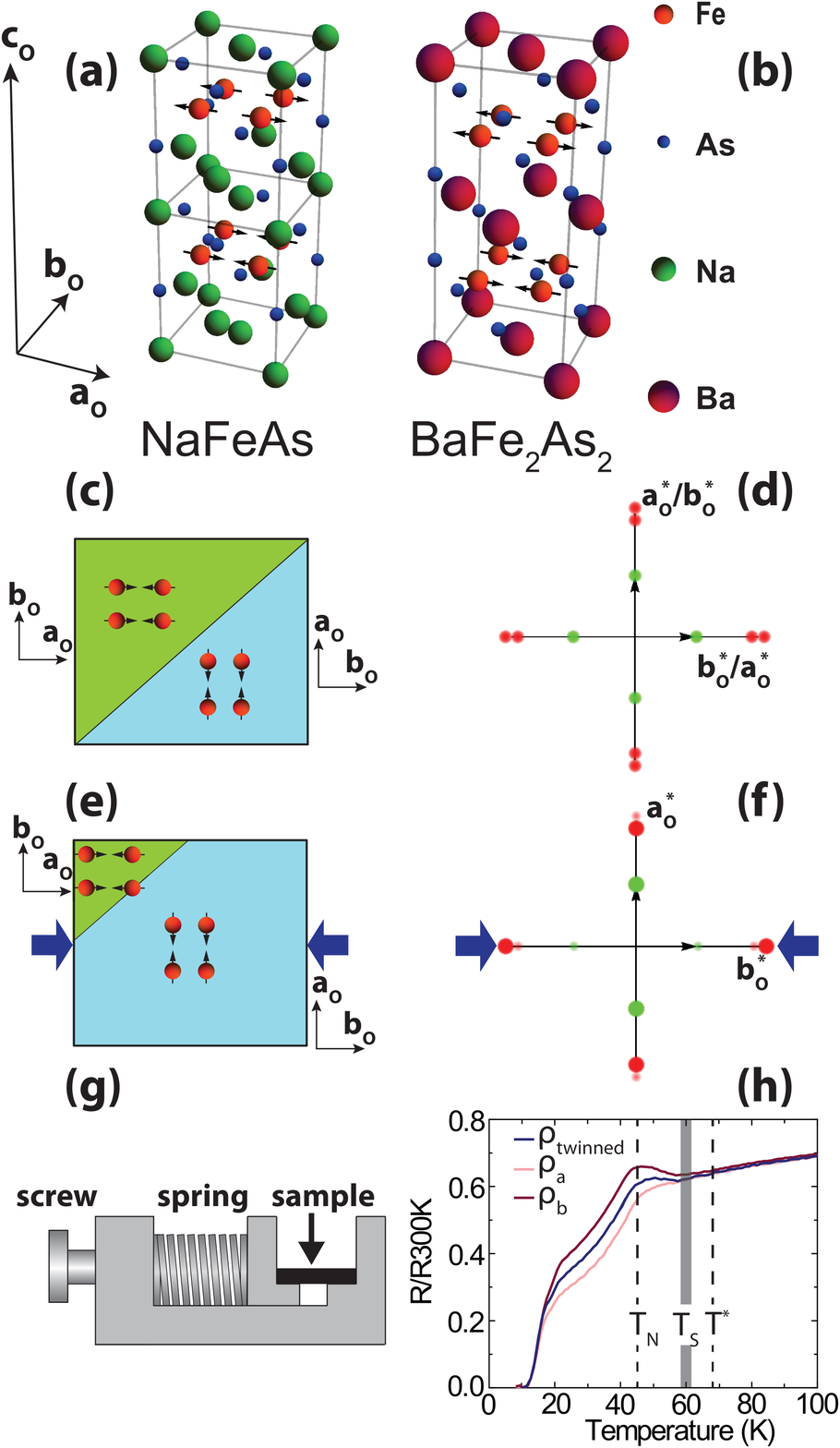}
\caption{
(Color online) The lattice and magnetic structures of (a) NaFeAs and (b) BaFe$_{2}$As$_{2}$.  While BaFe$_{2}$As$_{2}$ has the orthorhombic lattice and magnetic unit cells,  
NaFeAs consists of two orthorhombic chemical unit cells stacked along the $c$-axis. 
The real space schematics of a (c)twinned crystal and a (e) detwinned crystal, the two sets of domains have the same population for the twinned crystal whereas one set of domains dominates in the detwinned crystal. In reciprocal space, the magnetic and structural peaks corresponding to the two sets of domains have equal intensities for a (d) twinned crystal, while for a (f) detwinned crystal 
the dominant set of domains is enhanced while peaks corresponding to the minority set of domains have diminished intensities. Green spheres represent the magnetic $(1,0,L)$ peak and its equivalent points, red spheres represent the structural $(2,0,0)$ peak and its equivalent points. The blue arrows in (e) and (f) represent applied uniaxial pressure. (g) Schematic of the pressure device used in this work. Springs of known 
force constants and area of the sample edge were used to estimate the applied pressures. (h) Resistivity of twinned and detwinned NaFeAs. The dashed lines represent $T_{N}$ determined from neutron scattering and $T^\ast$ the onset temperature of resistivity anisotropy, the shaded region represents the temperature range of $T_{s}$ under different pressures found from neutron scattering results.
}
\end{figure}

Figures 1(a) and 1(b) show the schematic lattice and magnetic structures of NaFeAs and
BaFe$_2$As$_2$, respectively \cite{dai}.   On cooling from the high-temperature tetragonal state, 
NaFeAs exhibits a tetragonal-to-orthorhombic structural transition at $T_s\approx 58$ K
and then a paramagnetic-to-antiferromagnetic transition at $T_N\approx 47$ K \cite{slli}. For comparison, 
$T_s$ and $T_N$ in 
BaFe$_2$As$_2$ occur almost simultaneously below about $138$ K \cite{qhunag,mgkim}.  In the absence of 
uniaxial pressure, the low-temperature magnetic and crystal structures 
 have equally populated twinned domains with mixed AF orthorhombic states as shown in Fig. 1(c).  Figure 1(d) shows the $[H,K]$ 
 plane of the reciprocal space where the AF and crystalline lattice Bragg peaks 
 for a twinned sample are seen at $(\pm1,0)$/$(0,\pm1)$ and $(\pm2,0)$/$(0,\pm2)$ positions, respectively.  Upon applying uniaxial pressure along the orthorhombic
 $a_o/b_o$ direction \cite{fisher}, a single domain with sufficient large  size can be achieved  [Fig. 1(e)],
 the resulting AF Bragg peaks now occurring predominantly at $(\pm1,0)$ positions [Fig. 1(f)].
 
 \begin{figure}[t]
\includegraphics[scale=.4]{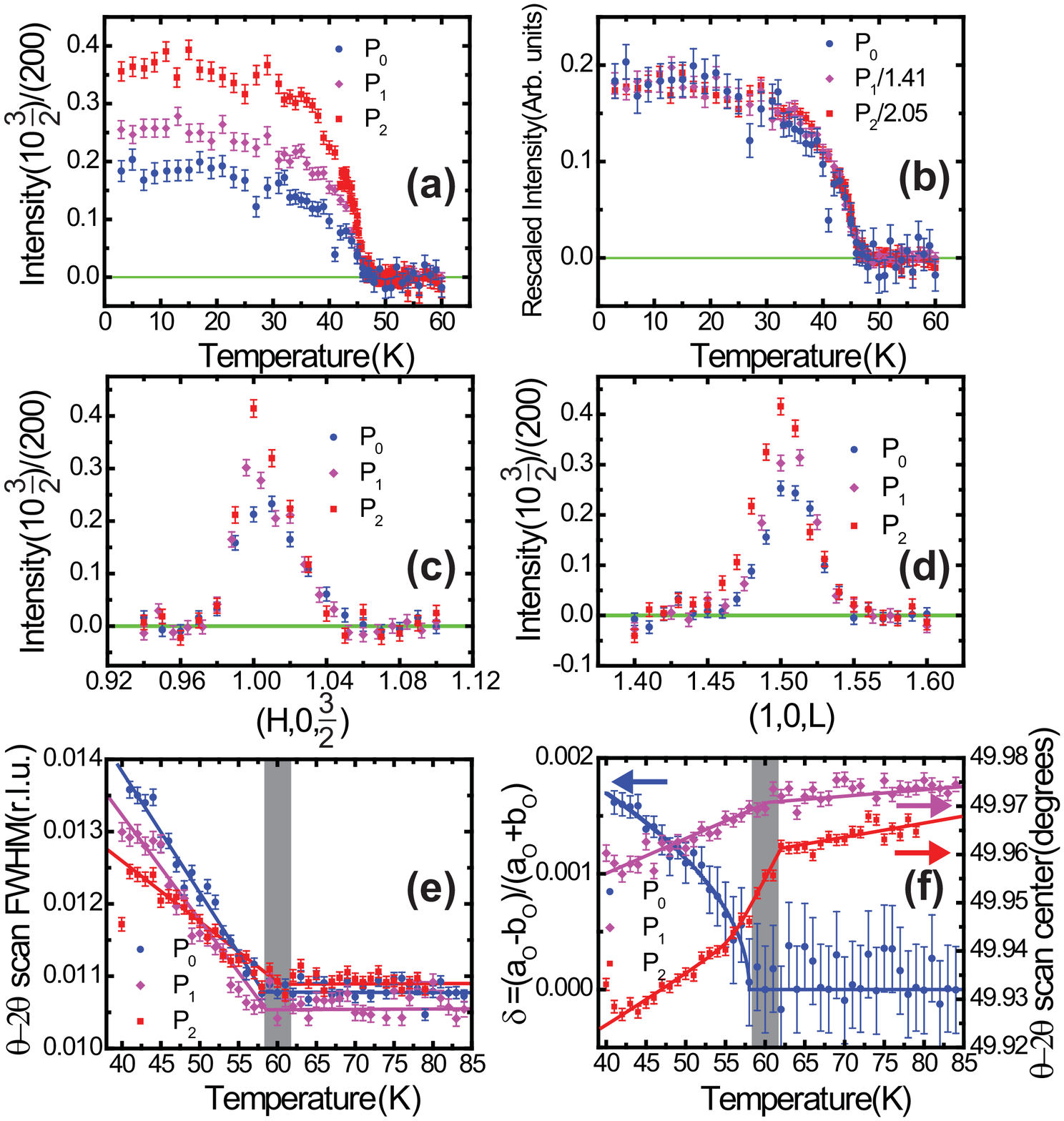}
\caption{(Color online) Elastic neutron scattering measurements on NaFeAs under ambient condition and with applied uniaxial pressures. $P_{0}$, $P_{1}$ and $P_{2}$ represent ambient condition,  $\approx$7 MPa applied uniaxial pressure and $\approx$15 MPa applied uniaxial pressure, respectively. 
The applied pressures are estimated from the changes in length of the spring and the known spring constant.
(a) Background subtracted magnetic order parameters measured at the $(1,0,1.5)$ peak normalized to the $(2,0,0)$ structural peak. (b) Magnetic order parameters normalized at base temperature
show within statistics (uncertainties represent one standard deviation) of the current measurement that uniaxial pressure does not affect its shape. (c) Background-subtracted $[H,0,1.5]$ scans for $(1,0,1.5)$ measured at 2.5 K normalized to $(2,0,0)$. (d) Background-subtracted $[1,0,L]$ scans for $(1,0,1.5)$ measured at 2.5 K normalized to $(2,0,0)$. (e) Full width half maximum (FWHM) of $\theta$-2$\theta$ scans at $(2,0,0)$ fit with a single Gaussian as a function of temperature. The solid lines are guides to the eye and the shaded region is the range of $T_{s}$ determined from panel (f). (f) Orthorhombicity of NaFeAs under ambient conditions determined from fits by two Gaussians of equal intensities and centers of $\theta$-2$\theta$ scans found from fitting a single Gaussian as a function of temperature. The solid lines are guides to the eye and the shaded region is the temperature range for $T_{s}$.
 }
 \end{figure}

We prepared high quality single crystals of NaFeAs, as well as as-grown and annealed BaFe$_2$As$_2$ crystals,
using the self-flux method.  The samples were cut to squared shapes
along the $a_o/b_o$ directions of the orthorhombic structure and fit into the aluminum-based
detwinning devices for both transport and neutron scattering experiments 
[Fig. 1(g)].  Figure 1(h) shows the comparison of 
transport measurements for both twinned and detwinned NaFeAs. Consistent with earlier measurements \cite{yzhang}, we find clear resistivity anisotropy below about $T^\ast\approx 70$ K.
 Our neutron scattering experiments were
carried out using the BT-7 triple-axis spectrometer at NIST center for 
neutron research (NCNR) \cite{lynn}, and 
HB-1A at High-Flux-Isotope-Reactor (HFIR), Oak Ridge National Laboratory.
For HB-1A measurements on BaFe$_2$As$_2$, the collimations are 
$48^\prime$-$48^\prime$-sample-$40^\prime$-$68^\prime$.  The magnetic measurements for NaFeAs were carried out
on BT-7 with open-$50^\prime$-sample-$50^\prime$-$120^\prime$ with $E_f=14.7$ meV.  
To separate the $(2,0,0)/(0,2,0)$ nuclear 
Bragg peaks in the orthorhombic state of a twinned sample [Fig. 1(d)], we used tight collimation
of $10^\prime$-$10^\prime$-sample-$10^\prime$-$25^\prime$ on BT-7 with $E_f=14.7$ meV.
For NaFeAs, the lattice parameters are $a_o=5.589$, $b_o= 5.569$ and $c=6.991$ \AA\ \cite{slli}.
BaFe$_2$As$_2$ has lattice parameters $a_o\approx b_o\approx 5.595$ \AA\ and $c=12.92$ \AA\ \cite{qhunag}.
The wave vector \textbf{Q} in three-dimensional reciprocal space in \AA$^{-1}$ is defined as 
${\bf Q}=H {\bf a_o^\ast}+K{\bf b_o^\ast}+L{\bf c^\ast}$, where $H$, $K$, and $L$ are Miller indices and
${\bf a_o^\ast}=\hat{{\bf a}}_o2\pi/a_o, {\bf b_o^\ast}=\hat{{\bf b}}_o 2\pi/b_o, {\bf c^\ast}=\hat{{\bf c}}2\pi/c$
are reciprocal lattice units (rlu). 
We aligned the crystals in the $[H,0,0] \times [0,0,L]$ scattering plane, where 
AF Bragg peaks occur at $(\pm1,0,L)$ with $L=\pm 0.5,\pm1.5,\cdots$ for NaFeAs \cite{slli} 
and $L=\pm 1,\pm 3,\cdots$ for BaFe$_2$As$_2$ \cite{qhunag}.
If the twinned domains are equally populated in the zero pressure state, AF Bragg peak intensity at
$(\pm1,0,L)$ should be the same as that at $(0,\pm1,L)$.  On the other hand, if uniaxial pressure completely detwinnes the sample, the magnetic scattering intensity at $(\pm1,0,L)$ in the detwinned state should increase by a factor of two compared with the twinned state.

The detwinning device we used is shown in Fig. 1(g). By knowing the 
compressibility of the spring and the area of the sample, we can estimate the applied uniaxial pressure.
For NaFeAs, we have $P_0=0$, $P_1\approx 7$ MPa, and  $P_2\approx 15$ MPa.  The applied uniaxial pressures are $P_1\approx 7$ and  $P_1\approx 6$ MPa for the as-grown and annealed
BaFe$_2$As$_2$ crystals, respectively.  We first describe the effect of uniaxial pressure on NaFeAs.  For our measurements, we always apply the pressure at room temperature, and measure the 
$(1,0,1.5)$ magnetic and $(2,0,0)$ nuclear Bragg peaks.  Figure 2(a) shows 
the
temperature dependence of the magnetic $(1,0,1.5)$ peak intensity normalized to the 
$(2,0,0)$ nuclear peak.
At zero pressure, we see a clear magnetic intensity increase below $T_N=47$ K.  On increasing to $P_1$ 
and then to $P_2$, we see that 
the magnetic scattering intensity almost doubles, suggesting that the uniaxial pressure has indeed detwinned the sample.
However, the N$\rm \acute{e}$el temperatures of the system remain unchanged at $T_N=47$ K within the errors of our measurements.  The normalized magnetic order parameter in Fig. 2(b)
shows almost identical behavior for $P_0$, $P_1$, and $P_2$, thus confirming that the uniaxial pressure needed to detwin NaFeAs has no measurable impact on $T_N$.
Figures 2(c) and 2(d) show wave vector scans along the $[H,0,1.5]$ and $[1,0,L]$ directions at $P_0$, $P_1$, and $P_2$.  Consistent with the order parameter data in Fig. 2(a), the effect of
uniaxial pressure is to increase the intensity of the AF Bragg peak $(1,0,1.5)$.  To probe the effect of uniaxial pressure on the tetragonal-to-orthorhombic lattice distortion temperature 
$T_s$, we studied the temperature dependence of the lattice orthorhombicity on the $(2,0,0)/(0,2,0)$ nuclear Bragg peaks
using tight collimations.  The full-width-half-maximum (FWHM) of $(2,0,0)/(0,2,0)$ shows a clear increase below
$T_s=58$ K for $P_0$, $P_1$, and $P_2$, and thus suggests that the applied uniaxial pressure also has only a small impact on $T_s$ [Fig. 2(e)].  Figure 2(f) shows the temperature dependence of the lattice orthorhombicity
$\delta=(a_o-b_o)/(a_o+b_o)$ at zero pressure and its comparison with the centers 
of $\theta$-2$\theta$ scans.  We can see a small ($\sim$4 K) increase in $T_s$ when NaFeAs is 
detwinned [Fig. 2(f)].

\begin{figure}[t]
\includegraphics[scale=.4]{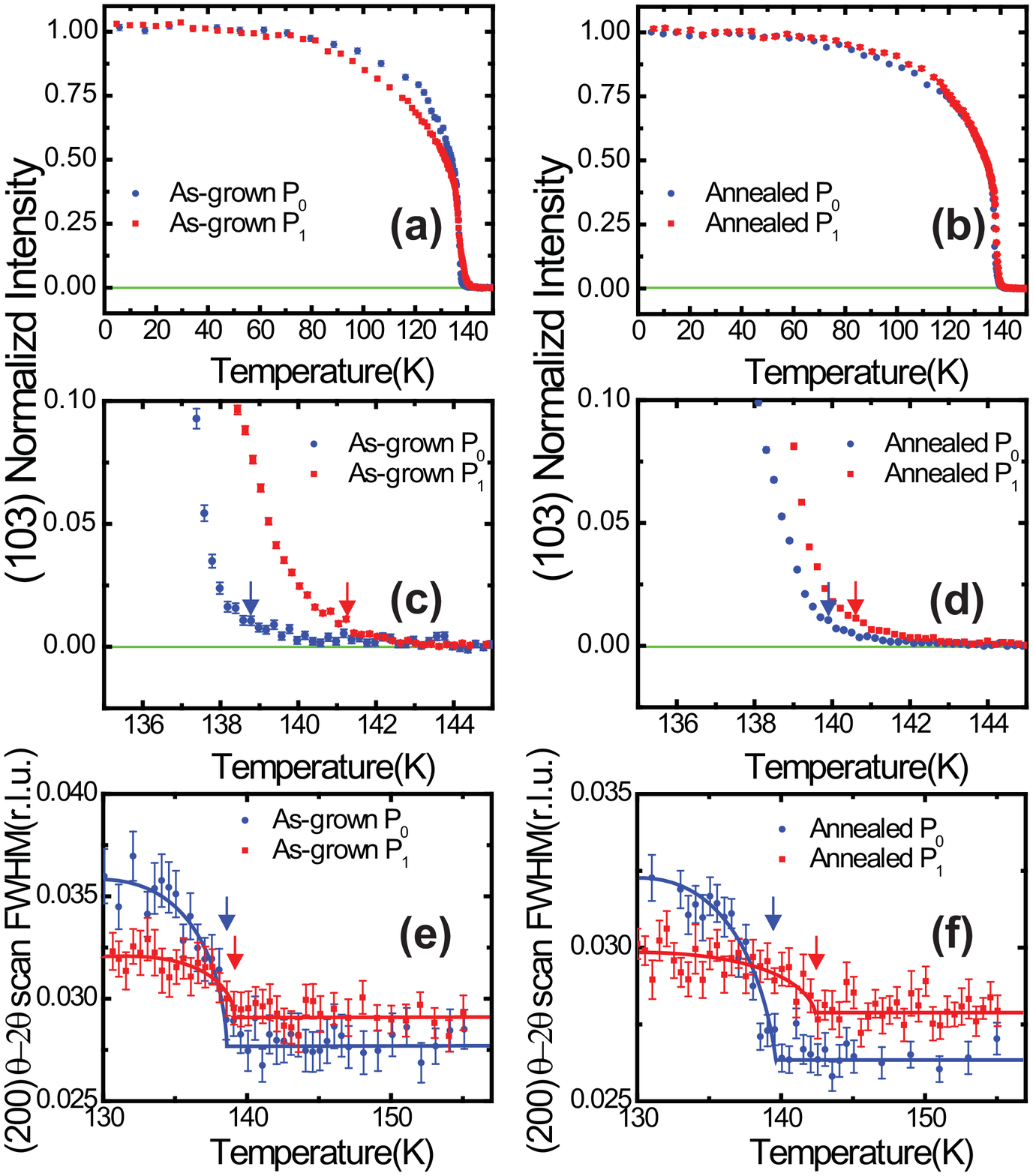}
\caption{  (Color online) Elastic neutron scattering measurements on as-grown and annealed BaFe$_{2}$As$_{2}$ under ambient conditions ($P_0$) and with applied uniaxial pressure ($P_{1}$ is $\sim$7 MPa for the as-grown sample and $\sim$6 MPa for the annealed sample). The background subtracted and normalized magnetic order parameters measured at $(1,0,3)$ for (a) as-grown and (b) annealed BaFe$_{2}$As$_{2}$ under ambient condition and with applied uniaxial pressure. (c) and (d) show expanded 
plots of the magnetic order parameter near the magnetic transition temperature. The arrows indicate temperatures at which the intensity reaches 1\% of the intensity at 2 K. Full width half maximum(FWHM) of $\theta$-2$\theta$ scans at $(2,0,0)/(0,2,0)$ for the (e) as-grown sample and the (f) annealed sample fit with single Gaussians as a function of temperature.
 }
\end{figure}

Having established that the uniaxial pressure needed to detwin NaFeAs has only a small impact on $T_N$ and $T_s$, we investigate the effect of uniaxial pressure on 
$T_N$ and $T_s$ in as-grown and annealed BaFe$_2$As$_2$.  From previous work on as-grown BaFe$_2$As$_2$, we know that the uniaxial pressure necessary to detwin the
sample will also increase the onset of $T_N$ and $T_s$ by $\sim$12 K \cite{dhital}.  On the other hand, transport measurements on as-grown and annealed BaFe$_2$As$_2$ suggest 
that the large resistivity anisotropy in detwinned as-grown samples is due to disorder in these materials 
and annealing significantly reduces the resistivity anisotropy \cite{ishida}.  To determine how uniaxial pressure affects as-grown and annealed BaFe$_2$As$_2$, we prepared annealed samples following Ref. \cite{ishida}, and carried out $T_N$ and $T_s$ measurements on HB-1A.
Figure 3(a,c) compares the low-temperature normalized AF $(1,0,3)$ Bragg peak intensities at $P_0=0$ and $P_1=7$ MPa for the as-grown  
BaFe$_2$As$_2$.  While the overall magnetic intensity behaves similarly with and without uniaxial pressure, we see a clear increase in $T_N$ from  
$\sim$139 K at $P_0=0$ to $\sim$141 K at $P_1=7$ MPa, consistent with the earlier work \cite{dhital}.  For the annealed BaFe$_2$As$_2$, similar measurements showed 
almost identical magnetic order parameters [Fig. 3(b,d)] and
a smaller shift in $T_N$ from $\sim$140 K at $P_0=0$ to $\sim$141 K at $P_1=6$ MPa.
Figure 3(e) plots the temperature dependence of the FWHM of the nuclear $(2,0,0)/(0,2,0)$ 
Bragg peak for the as-grown BaFe$_2$As$_2$.  
At $P_0=0$, the peak width increases abruptly below $T_s\approx 140$ K, reflecting the fact that a twinned orthorhombic crystal
has slightly different lattice parameters for $(2,0,0)$ and $(0,2,0)$.  $P_1=7$ MPa uniaxial pressure clearly increases the onset $T_N$ as shown in Fig. 3(c), while the $T_s$ under pressure 
only increases marginally to $T_s\approx 141$ K [Fig. 3(e)].  Similarly, we find that the uniaxial pressure of $P_1=6$ MPa on annealed BaFe$_2$As$_2$ only increases $T_s$ from $\sim$140 K 
to $\sim$143 K [Fig. 3(f)].

From the experimental data discussed above, it is clear that the uniaxial pressure necessary to detwin NaFeAs and annealed BaFe$_2$As$_2$ has 
limited impact on $T_N$ and $T_s$.  Theoretically, it has been argued that a small uniaxial strain of magnitude $A_0$ should induce an increase  
in the magnetic ordering temperature $\Delta T_N=\left|A_0\right|^{1/\gamma}$ if $T_s=T_N$, and $\Delta T_N=
(T_s-T_N)^{-\gamma}\left|A_0\right|$ if $T_s>T_N$, where the susceptibility exponent $\gamma=2+O(1/N)$ (with $N=3$ corresponding to the physically relevant Heisenberg case) \cite{jphu12}.  The structural transition temperature is also expected to increase on a scale of $\Delta T_s\sim \left|A_0\right|^x$, where for $N\rightarrow\infty$, $x=1+O(1/N)$ \cite{jphu12}.  
 Comparing with the nearly simultaneous structural and magnetic phase transitions in BaFe$_2$As$_2$ \cite{qhunag,mgkim}, the structural and magnetic phase
transitions in NaFeAs are separated by $T_s-T_N\approx 12$ K [Fig. 1(h)] \cite{slli}.  Within the spin nematic phase scenario 
\cite{jphu}, this arises because the 
$c$-axis magnetic exchange coupling in NaFeAs is much smaller than that of BaFe$_2$As$_2$ \cite{lharriger09,jtpark}. 
As a consequence, the shift of the N${\rm \acute{e}}$el temperature $\Delta T_N=
(T_s-T_N)^{-\gamma}\left|A_0\right|\approx \left|A_0\right|/144$ in NaFeAs should be much smaller than that ($\Delta T_N=\left|A_0\right|^{0.5}$) in BaFe$_2$As$_2$,
while the changes in structural transition temperatures ($\Delta T_s$'s) should be similar for both materials.  Indeed, while uniaxial strain seems to have some small effect on $T_s$ for
both NaFeAs [Fig. 1(f)] and BaFe$_2$As$_2$ [Figs. 3(e) and 3(f)], it has virtually no effect on $\Delta T_N$ for NaFeAs.  This means that the resistivity anisotropy seen in NaFeAs above $T_N$ and $T_s$ in Fig. 1(h)  
cannot be due to the orthorhombic lattice structure or collinear AF order.  Since NaFeAs also does not have Co as a source for anisotropic impurity scattering \cite{ishida,allan} and
the Na deficiency out of the FeAs plane is not expected to 
affect the transport measurements \cite{xhchen}, the resistivity anisotropy above $T_s$ in NaFeAs must be an intrinsic property of the paramagnetic tetragonal phase under uniaxial strain.  These results, together with the uniaxial pressure effect on as-grown and annealed BaFe$_2$As$_2$, suggest that the resistivity anisotropy in iron pnictide parent compounds cannot arise from strain-induced shift in $T_N$ and $T_s$.

In summary, we have shown that the uniaxial pressure needed to detwin NaFeAs and annealed BaFe$_2$As$_2$ 
has a very small effect on magnetic and structural phase transitions of these materials, while
transport measurements on identical materials reveal clear resistivity anisotropy above $T_s$.
We conclude then that the resistivity anisotropy is an intrinsic property in the uniaxial-strained paramagnetic tetragonal phase of NaFeAs and BaFe$_2$As$_2$.

We thank Jiangping Hu and Weicheng Lv for helpful discussions.
The single crystal growth and neutron scattering work at UTK is supported by
the US DOE BES No. DE-FG02-05ER46202. Work at the
IOP, CAS is supported by the MOST of China 973 program
(2012CB821400). The research at HFIR, ORNL 
was sponsored by the US DOE
Scientific User Facilities Division, Materials Sciences and Engineering Division, and BES.

\end{document}